\documentclass[fleqn,usenatbib]{mnras}

\usepackage{newtxtext,newtxmath}

\usepackage[T1]{fontenc}
\usepackage{ae,aecompl}

\usepackage{graphicx}	
\usepackage{dblfloatfix}
\usepackage{amsmath}	
\usepackage{threeparttable}
\usepackage{subcaption}
\usepackage{caption}
\usepackage{color}
\usepackage[dvipsnames]{xcolor}
\usepackage[normalem]{ulem}
\usepackage{booktabs}

\newcommand{\nicer}{\textit{NICER}}

\newcommand{\hstlong}{\textit{Hubble Space Telescope}}

\newcommand{\chandra}{\textit{Chandra}}
\newcommand{\gaia}{\textit{Gaia}}

\newcommand{\kteff}{\mbox{$kT_{\rm eff}$}}
\newcommand{\rinfty}{\mbox{$R_{\infty}$}}
\newcommand{\rns}{\mbox{$R_{\rm NS}$}}
\newcommand{\mns}{\mbox{$M_{\rm NS}$}}

\newcommand{\chisq}{\mbox{$\chi^2$}}
\newcommand{\chisqnu}{\mbox{$\chi^2_\nu$}}

\newcommand{\Chisq}[3]{$\chi^2_\nu$/dof (prob.) = {#1}/{#2} (#3)}

\newcommand{\halpha}{\mbox{H$\alpha$}}

\newcommand{\simlt}{\mathrel{\hbox{\rlap{\hbox{\lower4pt\hbox{$\sim$}}}\hbox{$<$}}}}
\newcommand{\simgt}{\mathrel{\hbox{\rlap{\hbox{\lower4pt\hbox{$\sim$}}}\hbox{$>$}}}}

\newcommand{\tee}[1]{\mbox{$\times 10^{#1}$}}
\newcommand{\ud}[2]{\mbox{\raisebox{0.4ex}{$\scriptstyle^{ + #1}_{- #2}$}}}
\newcommand{\ppm}{\mbox{$\pm$}}

\newcommand{\msun}{\mbox{$\,M_\odot$}}

\newcommand{\km}{\hbox{$\,{\rm km}$}}

\newcommand{\kpc}{\mbox{$\,{\rm kpc}$}}

\newcommand{\persec}{\mbox{$\,{\rm s^{-1}}$}}

\newcommand{\percmsq}{\mbox{$\,{\rm cm^{-2}}$}}

\newcommand{\cgsflux}{\mbox{$\,{\rm erg\,\percmsq\,\persec}$}}

\newcommand{\OmCen}{\mbox{$\omega$\,Cen}}

\title[M-R constraints on qLMXB in M30]{Spectral analysis of the quiescent low-mass X-ray binary in the globular cluster M30}

\author[C. S. Echibur\'u et al.]{
C. S. Echibur\'u,$^{1,2}$\thanks{E-mail: constanza.echiburu@mail.mcgill.ca},
S. Guillot$^{3}$\thanks{E-mail: sebastien.guillot@irap.omp.eu},
Y. Zhao$^{4}$,
C. O. Heinke$^{4}$,
F. \"{O}zel$^{5}$,
and 
\newauthor
N. A. Webb$^{3}$
\\
$^{1}$McGill University, Department of Physics, McGill Space Institute, Montr\'eal, QC, H3A 2A7, Canada\\
$^{2}$Instituto de Astrof\'isica, Pontificia Universidad Cat\'olica de Chile, Av. Vicu\~na Mackenna 4860, 7820436 Macul, Santiago\\
$^{3}$IRAP, Universit\'e de Toulouse, CNRS, CNES, Toulouse, France\\
$^{4}$Department of Physics, University of Alberta, CCIS 4-183, Edmonton, AB T6G 2E1, Canada\\
$^{5}$Department of Astronomy, University of Arizona, 933 N. Cherry Ave., Tucson, AZ 85721
}

\date{Accepted XXX. Received YYY; in original form ZZZ}

\pubyear{2019}

\begin{document}
\label{firstpage}
\pagerange{\pageref{firstpage}--\pageref{lastpage}}
\maketitle

\begin{abstract}
We present a recent \emph{Chandra} observation of the quiescent low-mass X-ray binary containing a neutron star, located in the globular cluster M30. We fit the thermal emission from the neutron star to extract its mass and radius. We find no evidence of flux variability between the two observations taken in 2001 and 2017, nor between individual 2017 observations, so we analyse them together to increase the signal to noise. We perform simultaneous spectral fits using standard light-element composition atmosphere models (hydrogen or helium), including absorption by the interstellar medium, correction for pile-up of X-ray photons on the detector, and a power-law for count excesses at high photon energy. Using a Markov-chain Monte Carlo approach, we extract  mass and radius credible intervals for both chemical compositions of the atmosphere: $\rns=7.94\ud{0.76}{1.21}$~km and $\mns<1.19$~M$_{\odot}$ assuming pure hydrogen, and $\rns=10.50\ud{2.88}{2.03}$~km and $\mns<1.78$~M$_{\odot}$ for helium, where the uncertainties represent the 90\% credible regions.  For H, the small radius is difficult to reconcile with most current nuclear physics models (especially for nucleonic equations of state) and with other measurements of neutron star radii, with recent preferred values generally in the 11--14~km range. Whereas for He, the measured radius is consistent with this range. We discuss possible sources of systematic uncertainty that may result in an underestimation of the radius, identifying the presence of surface temperature inhomogeneities as the most relevant bias. According to this, we conclude that either the atmosphere is composed of He, or it is a H atmosphere with a significant contribution of hot spots to the observed radiation.

\end{abstract}





\section{Introduction}
\label{sec:intro}

In the complex interior of neutron stars (NSs), and particularly in their cores, matter can be found at extreme densities, higher than $2.8 \times 10^{14}$ g cm$^{-3}$, the nuclear matter density. Since those conditions cannot be reproduced on Earth, NSs provide a unique laboratory to study the behaviour of matter at such high densities. The composition and particle interactions are described by an equation of state (EoS), a relation between pressure and energy density at any point inside the star. For a given EoS, solving the Tolman-Oppenheimer-Volkoff equations of stellar structure \citep{1939PhRv...55..364T,1939PhRv...55..374O} for a range of central densities results in a unique mass-radius (\mns-\rns) relation for NSs. Therefore, determining $\mns$ or $\rns$ by observational methods \citep[see][for a recent review]{2016ARA&A..54..401O} is crucial to discriminate between EoSs proposed by nuclear theory. These are tested by their capability to reproduce the observed macroscopic properties of NSs. Unfortunately, a simultaneous measurement of $\mns$ and $\rns$ is difficult to achieve.

The recently launched Neutron star Interior Composition Explorer (\nicer, \citealt{gendreau17}), a soft X-ray telescope onboard the International Space Station, was designed to address this problem. \nicer\ is characterising the pulse profiles of rotation-powered millisecond pulsars, with a goal of measuring their masses and radii to an accuracy of 5-10\% (see \citealt{2019arXiv190407012W}, and \citealt{2019ApJ...887L..21R,2019ApJ...887L..24M} for the measurements of the pulsar PSR~J0030+0451). Other mass measurements have been obtained by means of X-ray and optical observations of NSs with high and low-mass companions \citep[e.g.,][]{2011ApJ...730...25R}, but the most precise masses have generally been extracted from radio timing observations of pulsars in binary systems \citep[e.g.,][]{2012MNRAS.423.3316A}. Radius measurements, on the other hand, are more difficult to achieve, since existing methods suffer from significant systematic errors \citep[see e.g.,][]{miller13,2014MNRAS.444..443H,2016ApJ...831..184B}.

One promising method to constrain NS radii consists of the study of low-mass X-ray binaries (LMXBs) hosting NSs, during quiescence (qLMXBs). Accretion onto compact objects has been observed to be a transient process. When accretion activity increases, X-ray luminosities may rise to $\sim10^{36-38}$~erg~s$^{-1}$ for weeks to months periods. In these outburst episodes, accreted matter accumulates on the NS surface and then form layers of increasing element mass (as a function of depth) due to the large surface gravity. Because of the continuous accretion during outbursts, these accreted layers sink deeper inside the NS crust under the weight of the newer ones, i.e., to high densities, which induces nuclear reactions. The released energy is deposited in the crust and core, heating them up, as explained by \emph{deep crustal heating} \citep{1998ApJ...504L..95B}. 

In contrast, the quiescent emission observed after active periods of accretion\footnote{In quiescence, accretion ceases or continues at a very low level $\lesssim$ 10$^{-12}$ M$_{\odot}$ yr$^{-1}$.}, has typical luminosities in the $10^{32-33}$~erg~s$^{-1}$ range, and can last for periods of years to decades. The origin of this thermal emission is the heat released from the crust and reprocessed by the atmosphere, which then emerges as a $\sim 10^{6}$~K thermal X-ray spectrum at the surface of the NS.

Since NSs in qLMXBs have been accreting from low-mass stars, and because sedimentation occurs on short timescales \citep{alcock80,bildsten92}, their atmosphere is likely to be composed of hydrogen \citep{2002ApJ...574..920B}, although this depends on the nature of the donor star which could be H-depleted. On the surface of NSs in qLMXBs, hydrogen is fully ionized due to the high temperatures of $\sim$10$^6$ K \citep{1996A&A...315..141Z}, facilitating free-free absorption. In these processes, the opacity is given by $\kappa \propto E^{-3}$, where $E$ is the energy of the photon, which causes the atmosphere to be less opaque to photons with larger energies. As a result, photons coming from deeper and hotter layers propagate more easily through the atmosphere, shifting the peak of the spectrum to higher energies (by a factor 1.6--1.8) compared to a black body spectrum with the same effective temperature \citep{1996A&A...315..141Z,1996ApJ...461..327R}. 

An atmosphere model allows us to extract the radius measured at infinity $R_{\infty} = \rns(1+z) = \rns/\sqrt{1-2G\mns/\rns c^2}$ through spectral fitting, but this requires high-quality spectra and, more importantly, knowing the distance to the source. It is known that globular clusters (GCs) host numerous binary systems, due to their high stellar densities and the resulting stellar interactions \citep{2006csxs.book..341V}. If the qLMXB is located in a GC with a well constrained distance\footnote{Typical GC distance measurements have $\sim$5--10\% uncertainties \citep[e.g.,][]{2017ApJ...838..162O}.}, this provides an independent measurement and reduces the uncertainties in the spectral modeling. Another advantage of qLMXBs is that, since they are thought to be the progenitors of millisecond pulsars \citep[e.g.,][]{1982Natur.300..728A,1982A&A...108..143S,1991PhR...203....1B,2006csxs.book..623T}, they are expected to have low magnetic fields, $\ll 10^{10}$~G, such that the opacities remain unaffected by the magnetic field, and so does the emergent spectrum \citep{1996A&A...315..141Z}.

Several qLMXBs have been identified in GCs and in the Galactic field (for some examples, see Table~4 in \citealt{2009MNRAS.392..665G} and references therein). While LMXBs in the field were detected following the onset of a bright accretion outburst, most qLMXBs in GCs, including all those with the highest flux at Earth, have not shown accretion activity\footnote{A few transients hosted in GCs were known to be very faint X-ray sources prior to their first observed outburst, but their quiescent spectrum was too faint to be well characterized (e.g., IGR~J17480--2446 in Terzan~5, \citealt{heinke06b,bordas10,pooley10}).}. Most of these sources have only been spectrally identified based on their similarities to field LMXBs, observed during quiescence (e.g., Cen~X-4 or Aql~X-1). Previous works have confirmed that H-atmosphere models accurately describe the spectra of qLMXBs, with radii in the range 10--15~km, as expected for NSs, either from single sources  \citep[e.g.,][]{2006ApJ...644.1090H,2007ApJ...671..727W,2011ApJ...732...88G,2014MNRAS.444..443H,2016ApJ...831..184B}, or from statistical analyses of multiple qLMXBs \citep[e.g.,][]{2013ApJ...772....7G,2014ApJ...796L...3G,lattimer14,guillot16b,2018MNRAS.476..421S}. However, in some cases the accreted material may not be hydrogen, but helium \citep[e.g.,][]{2012MNRAS.423.1556S,2013ApJ...764..145C,2014MNRAS.444..443H}. One way to circumvent this is to identify the nature of the donor star, i.e., to determine the nature of the material transferred onto the NS \citep[e.g., with the detection of a H$\alpha$ emission line, presumably originating in a faint accretion disk,][]{2004ApJ...613..512H}. The possibility of helium (or heavier element) atmospheres is well-founded on the existence of ultra-compact X-ray binaries (UCXB), with white dwarfs or helium dominated donors\footnote{However, only $10^{-20}\msun$ of hydrogen is necessary to produce an optically thick NS atmosphere \citep{2016ApJ...831..184B}, and the matter transferred from He rich donors may still possess enough H to dominate the NS atmosphere, after gravitationnal settling.} \cite[e.g.,][]{2009ApJ...699.1113Z,2010ApJ...712L..58A,2017A&A...598A..34S,2019ApJ...875...25C}. In fact, around 1/3 of the LMXBs in GCs with constraints on the companion nature, possess a white dwarf donor \citep{2014ApJ...780..127B}. Since NS He-atmosphere models have harder spectra than H-atmosphere models, using the incorrect composition for the observed thermal emission can result in biases of the inferred radii \citep{2012MNRAS.423.1556S,2014MNRAS.444..443H}.

The motivation for the present work is to provide updated $\mns$ and $\rns$ measurements of the NS in the qLMXB located in the core of the GC M30. This source (CXOGLB~J214022.13--231045.5, also named M30 A1) was originally reported by \cite{2007ApJ...657..286L} using a \emph{Chandra} observation obtained in 2001. In that work, the authors considered a H-atmosphere model corrected for pile-up (see description in Section~\ref{sec:analysis}). The results were rather unconstrained, with ranges: $7.0 \leq \rns \leq 19.2\km$ and $\mns\leq 2.8\msun$. Here, we present new results that include recent deep \emph{Chandra} observations from 2017, using both hydrogen and helium atmospheres. We suggest here a preference for a helium atmosphere composition, which results in a NS radius more compatible with the expected range (from observations of other NSs and from predictions of nuclear theory). However, if the atmosphere is composed of H, the presence of un-modeled hot spots on the surface of the NS could also explain our results.

The paper is organized as follows. In Section~\ref{sec:data} we describe the observations and data reduction. In Section~\ref{sec:analysis} we present our analyses prior to the spectral fitting, and our spectral fits. We discuss the results in Section~\ref{sec:discussion}. Finally, we present our conclusions in Section~\ref{sec:conclusion}.


\section{Observations and data reduction}\label{sec:data}

In this work we analyse nine \emph{Chandra} observations of the GC M30, separated in two sets. The first one corresponds to a 49~ks full-frame ACIS-S observation obtained in 2001 and reported in \cite{2007ApJ...657..286L}. In 2017, we obtained eight additional observations with the ACIS-S detector in 1/8 sub-array mode, totalling 281~ks of additional exposure designed to limit the effects of pile-up. Details for both sets of observations are given in Table~\ref{tab:data}. We performed the data reduction and analysis using {\sc ciao} v4.10 \citep{2006SPIE.6270E..1VF}, according to the steps suggested in the data preparation thread\footnote{http://cxc.harvard.edu/ciao/threads/data.html}. The level-1 data were first reprocessed with the latest calibration database CALDB v4.8.2, running the script \texttt{chandra\_repro}, which results in new level-2 event files used in the subsequent analysis.

\begin{figure*}
	\includegraphics[width=0.4\textwidth]{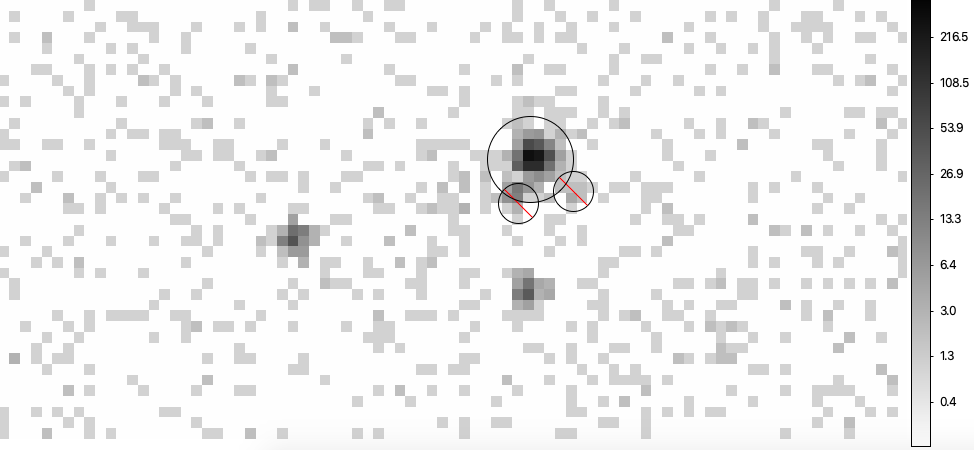}
	\includegraphics[width=0.4\textwidth]{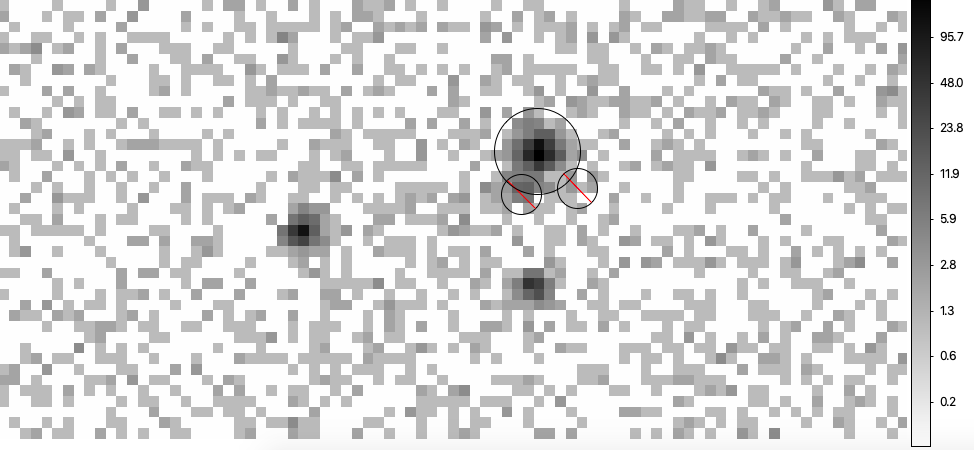}
    \caption{42$\arcsec \times$20$\arcsec$ Chandra X-ray images of the qLMXB in M30, in the 0.3--10 keV energy band. The figure shows the extraction regions for two of the observations: ObsID 2679 from 2001 (left) and ObsID 18997 from 2017 (right). The larger circle corresponds to the source region. The two objects located south of the qLMXB (likely to be cataclysmic variables, sources $A2$ and $A3$ in \citealt{2007ApJ...657..286L}), are excluded from the source region.}
    \label{fig:srcs}
\end{figure*}

The position of the qLMXB was first reported by \cite{2007ApJ...657..286L}, after astrometric correction with \hstlong\ images: $\alpha = 21^h40^m22^s.130 (1)$ and $\delta = -23^{\circ}10'45\farcs57(1)$. We extract the source spectra from circular regions of radius $2\arcsec$, which ensures that $\approx$97\% of the enclosed energy fraction at 1 keV is included\footnote{\textit{Chandra} Observatory Proposer Guide v21.0, Figure 6.10, 2018 December}. We used \texttt{wavdetect} to obtain the centroid position of the source on each observation, and the extraction is performed following the Step-by-Step guide in the data preparation guide, excluding the two objects south of the qLMXB (see Figure~\ref{fig:srcs}), considering a circular region of radius $0.9\arcsec$. These correspond to the sources $A2$ and $A3$ in \citealt{2007ApJ...657..286L}\footnote{The analyses of other cluster sources will be reported in Zhao et al. 2020, in preparation}. We estimated the number of counts from these two sources possibly contaminating the qLMXB region. For ObsID 2679, we found that 2.2 counts fall within this region, i.e., <0.2\% of the total count rate of the qLMXB ($\sim$1100 counts). The background spectra were extracted from a source-free $46.3\arcsec$ radius circular region near the qLMXB, on the same detector. Notice that the Auxiliary Response Files (ARFs) were corrected for the extraction region. We binned events in each spectrum using the \texttt{heasoft} tool \texttt{grppha} to 20 counts per bin, in the energy range 0.3--10~keV. Channels below 0.3~keV were ignored because \textit{Chandra}'s calibration is uncertain there, and above 10~keV since the sensitivity of the ACIS instrument decreases rapidly in that range. The instrumental calibration uncertainties were accounted for by adding 3\% systematic error. The eight resulting 2017 spectra are combined into a single spectrum using the {\sc ciao} tool \texttt{combine\_spectra}, and then binned following the procedure described before. Therefore, we analyse 2 spectra: one corresponding to the 2001 observation, and the other corresponding to the combined 8 spectra of 2017 observations.

\begin{table*}
\centering
\caption{Summary of the available \emph{Chandra} observations of M30.}
\label{tab:data}
\begin{threeparttable}
\begin{tabular}{cccccc} 
\hline
Obs. ID  & Starting Time & Exposure Time & Count Rate$^{\textrm{a}}$ & Detector$^{\textrm{b}}$ & Mode \\
& &(ks)&(s$^{-1}$)\\
\hline
2679	& 2001-11-19 02:55:12 & 49.43 &	0.02  &	ACIS-S FF  & FAINT \\
18997	& 2017-09-06 00:05:19 & 90.19 & 0.007 &	ACIS-S 1/8 & VFAINT\\
20725	& 2017-09-04 16:33:05 & 17.49 &	0.007 & ACIS-S 1/8 & VFAINT\\
20726	& 2017-09-10 02:09:13 & 19.21 & 0.007 &	ACIS-S 1/8 & VFAINT\\
20731	& 2017-09-16 18:04:17 & 23.99 & 0.009 & ACIS-S 1/8 & VFAINT\\
20732	& 2017-09-14 14:23:17 & 47.90 & 0.007 &	ACIS-S 1/8 & VFAINT\\
20792	& 2017-09-18 04:21:43 & 36.86 & 0.007 &	ACIS-S 1/8 & VFAINT\\
20795	& 2017-09-22 11:39:56 & 14.33 & 0.006 & ACIS-S 1/8 & VFAINT\\
20796	& 2017-09-23 06:09:30 & 30.68 & 0.007 &	ACIS-S 1/8 & VFAINT\\
\hline
\end{tabular}
\begin{tablenotes}
\item[a] \scriptsize The count rate difference is due to the decrease in the sensitivity of \emph{Chandra}-ACIS at low-energies, in the 16 years that separate the observations (\url{http://cxc.harvard.edu/ciao/why/acisqecontam_previous.html}).
\item[b] \scriptsize "FF" refers to full-frame mode and "1/8" refers to the 1/8 subarray mode.
\end{tablenotes}
\end{threeparttable}
\end{table*}

\section{Spectral analysis} \label{sec:analysis}

The spectral analysis is divided in three parts. First, since the two data sets were obtained 16 years apart, we search for possible variability of the flux. The second and third parts consist in the simultaneous spectral fits using H and He atmosphere models, respectively. All the spectral fits in the present work are performed with the X-ray analysis package {\sc xspec} v12.10.0c \citep{1996ASPC..101...17A} and using \chisq\ statistics. We set the 2001 data set as the spectral \texttt{group 1} and the 2017 data as the spectral \texttt{group 2}. In the following paragraphs we provide a brief description of each model used for the spectral fits.

For the H-atmosphere composition, we use the common model \texttt{nsatmos} \citep{2006ApJ...644.1090H}, and for the He-composition, we use the \texttt{nsx} model \citep{2009Natur.462...71H}. In both cases, the normalisation remains fixed to unity, i.e., assuming that the entire NS surface emits uniformly. We employ a distance $d=8.20$~kpc to M30 \citep{2017ApJ...838..162O}, the most recent value for this globular cluster. The absorption by the interstellar medium is accounted for with the \texttt{tbabs} model \citep{2000ApJ...542..914W}, with abundances set to {\tt wilm} \citep{2000ApJ...542..914W} and cross-sections set to {\tt vern} \citep{1996ApJ...465..487V} 

We also considered the effect of pile-up \citep{2001ApJ...562..575D}. This phenomenon occurs when two or more photons are registered on a single pixel within one readout time frame of the detector. As a consequence, the readout electronics interpret this event as one photon, with total energy equal to the sum of the energy of the incident photons. This causes a distortion in the shape of the spectrum, and since the inferred NS radius depends on the spectral shape, the effect of un-modeled pile-up could lead to biased radius measurements. The pile-up fractions for the qLMXB in M30 observations are low but not negligible: $\sim$2\% for 2001 data (with 0.062 counts per read-out frame), and $<1\%$ for 2017 data (0.006 counts per frame). These were obtained from the \emph{Chandra} proposal tool PIMMS\footnote{https://asc.harvard.edu/toolkit/pimms.jsp}.
Nevertheless, such small values can substantially impact the results. For instance, \cite{2016ApJ...831..184B} demonstrated that the inferred radius for the qLMXB X7 in 47Tuc may be underestimated by $\sim$10\% when pile-up effects, even with $\sim$1\% fraction, are neglected. Therefore, modelling pile-up even for a small fraction is fundamental to obtain reliable measurements of \mns\ and \rns. The \texttt{pileup} model provided in {\sc xspec} is fully described in \citet{2001ApJ...562..575D}. The parameters of this model component are set according to each observation and \emph{The Chandra ABC Guide To Pileup}\footnote{http://cxc.harvard.edu/ciao/download/doc/pileup\_abc.pdf}. We set a frame time of 3.14 s for \texttt{group 1} (corresponding to the full-frame mode readout time), and 0.941 s for \texttt{group 2} (for the 1/8-frame mode data), while the $\alpha$ parameter of both groups is left untied between the two, and free to vary (see Table \ref{tab:nsatmos_table}). The remaining parameters of the pile-up model are fixed to their default values.

Additionally, a high-energy excess has been observed in the tail of some LMXBs in quiescence \citep[e.g.,][]{1998A&ARv...8..279C,2002ApJ...577..346R,2004ApJ...601..474C,2011MNRAS.412.1409D,2014ApJ...780..127B}. Its origin is unclear and might be caused by residual accretion. We explore this possibility by adding a power-law spectral component, \texttt{pegpwrlw}, to the \texttt{nsatmos} or \texttt{nsx} model. To limit degeneracies between the thermal and non-thermal components during the spectral fits, we fix the photon index to 1.5, since the typical range observed for qLMXBs X-ray power spectrum is 1--2, as done in previous analyses of qLMXBs \citep{2013ApJ...772....7G,2016ApJ...831..184B}. The power-law energy range is set to its default value, i.e., 2-10 keV, and the power-law normalisation is left untied between the two groups.

Finally, to determine the best-fit parameters and their uncertainties, we employ the \texttt{chain} command in {\sc xspec} that performs Markov Chain Monte Carlo (MCMC) simulations to obtain the posterior distributions of all fitted parameters. The MCMC is initialised from the best-fit, with 200000 steps, 100 walkers, and burn-in corresponding to 25\% of the chain length. The priors are uniformly distributed within the hard limits of the \texttt{nsatmos} or \texttt{nsx} parameters. For \texttt{nsatmos}, these are 0.5--3.0 $M_\odot$ for \mns, 5--30 km for \rns, and 5--6.5 for $\log T_{\mathrm{eff}}/K$. The \texttt{nsx} model shares the same limits, except for $\log T_{\mathrm{eff}}/K$, covering the 5.5--6.7 range. All uncertainties reported hereafter are at the 90\% confidence level, or correspond to the 90\% credible interval when derived from the MCMC posterior distributions, unless specified otherwise.

\subsection{Flux variability} \label{sec:flux_comp}

The X-ray flux of qLMXBs is not expected to show variability during quiescence, unless an accretion episode happened recently\footnote{Typical returns to quiescence flux levels following an outburst can last several years \citep[e.g.,][for a short review]{2017JApA...38...49W}.}. However, such an event would likely be detected by all-sky monitors, as was the case for IGR~J17480--2446 in Terzan~5 \citep{bordas10,pooley10}.  We search for signs of flux variation over the 16 years between the two observations, and additionally, between the individual 2017 observations. The lack of variability would allow us to fit both spectra simultaneously with tied NS parameters. If the spectra show variability, it would however imply a change in temperature or in the normalisation of the thermal spectrum. That was observed in two qLMXBs (in NGC~6440 CX~1 and Terzan~5 CX~12,  \citealt{,2005ApJ...620..922C,2015MNRAS.449.1238W,2015MNRAS.452.3475B}), which showed marginal evidence of variations due to low-level accretion. 

We begin by testing for flux variations in timescales of years. For this exercise, the spectral model \texttt{tbabs*nsatmos} was convolved with \texttt{cflux}, a {\sc xspec} model used to estimate the flux of the other components, and their corresponding uncertainties. We fit the data of \texttt{group 1} (2001) and \texttt{group 2} (2017) with all parameters untied. The spectral fit results in fluxes $\log_{10}(F_{2001}/\cgsflux)=-13.12\ud{0.03}{0.03}$ for the 2001 data, and $\log_{10}(F_{2017}/\cgsflux)=-13.11\ud{0.03}{0.03}$ for the 2017 data. The temperatures were measured independently using the spectral model \texttt{pileup*tbabs*nsatmos} (see Section~\ref{sec:atmospheres}). We find $\kteff=104.23\ud{23.48}{14.26}$ eV for the 2001 data, and $\kteff=112.58\ud{25.06}{10.79}$ eV for the 2017 data.
For this spectral fit, we found a statistically acceptable result, with $\chi^{2}/{\rm dof}\equiv\chisqnu=0.97$, where dof is the number of degrees of freedom, corresponding to 101, and null hypothesis probability (nhp) equal to 0.58. Therefore, there is no evidence of flux variability of this source between 2001 and 2017.\\
Additionally, we tested an alternative method to search for flux variability. For this exercise we fit the model \texttt{pileup*tbabs*(const*nsatmos)}, fixing \texttt{const} to 1 in \texttt{group 2}, and with the remainder parameters free and tied between the two groups. Here, the \texttt{const} factor represents the relative normalisation of the two fluxes.
We found a value of the constant consistent with unity, 0.97\ud{0.11}{0.10}, were the errors correspond to the 99\% confidence estimated with \texttt{error} command. Thus, there are no signs of flux variability, leading to the same conclusion.

\begin{figure}
    \centering
	\includegraphics[width=0.45\textwidth]{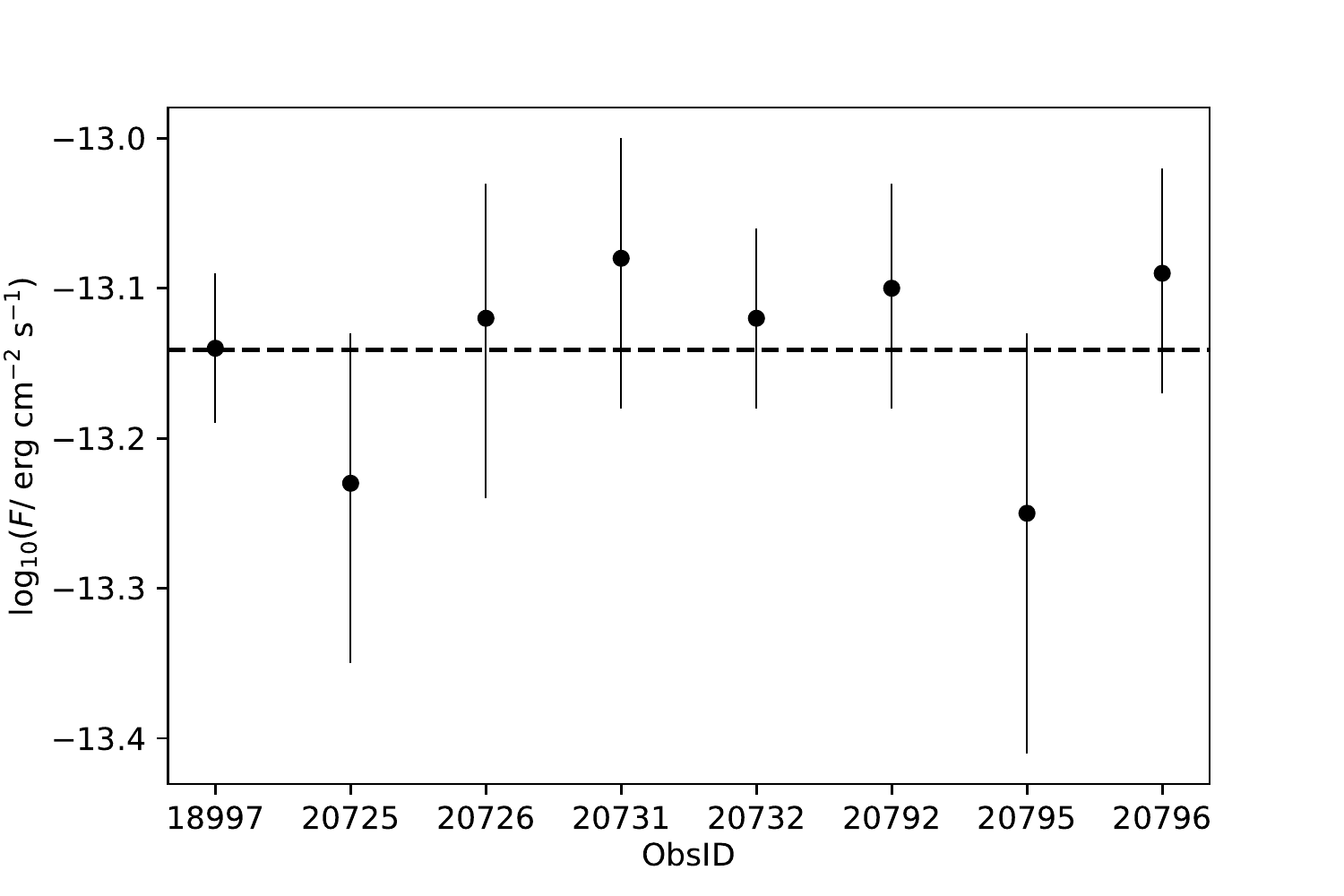}
    \caption{Flux estimation and 2$\sigma$ error bars for individual 2017 observations. The dashed line indicates the best-fit flux $\log_{10}(F/\cgsflux)=-13.14\ud{0.02}{0.02}$ (\chisqnu = 0.004) obtained from least-squares fitting method, where the error corresponds to the estimated standard error for the best-fit value.}
    \label{fig:fluxvar}
\end{figure}

The quiescent spectrum of LMXBs can also show short-term variability over timescales of days \citep[e.g.,][]{2018MNRAS.479.2777R}. Therefore, we search for flux variations within the individual observations taken in 2017. Due to the low number of counts in most 2017 observations (except ObsIDs 18997 and 20792), we rebin the data following the procedure described in Section \ref{sec:data} but using a minimum of 15 counts per bin. We fit the spectral model \texttt{tbabs*nsatmos} convolved with \texttt{cflux} to each observation, and computed the best-fit flux from MCMC runs. We do not include the \texttt{pileup} model, since 2017 data was obtained in 1/8-frame mode to reduce pile-up fraction, making this model difficult to constrain given the count rates shown in Table \ref{tab:data}. The results are shown in Figure \ref{fig:fluxvar} (see also Table \ref{tab:indv_fluxes}), and indicate the flux is constant within uncertainties, placing this source among the few steady qLMXB reliable for measuring radii \citep[e.g.,][]{2006ApJ...644.1090H,2012MNRAS.423.1556S,2013ApJ...772....7G}.

\subsection{Hydrogen and Helium atmospheres} 
\label{sec:atmospheres}
\begin{figure*}
    \centering
    \begin{subfigure}{0.495\textwidth}
        \includegraphics[width=\textwidth]{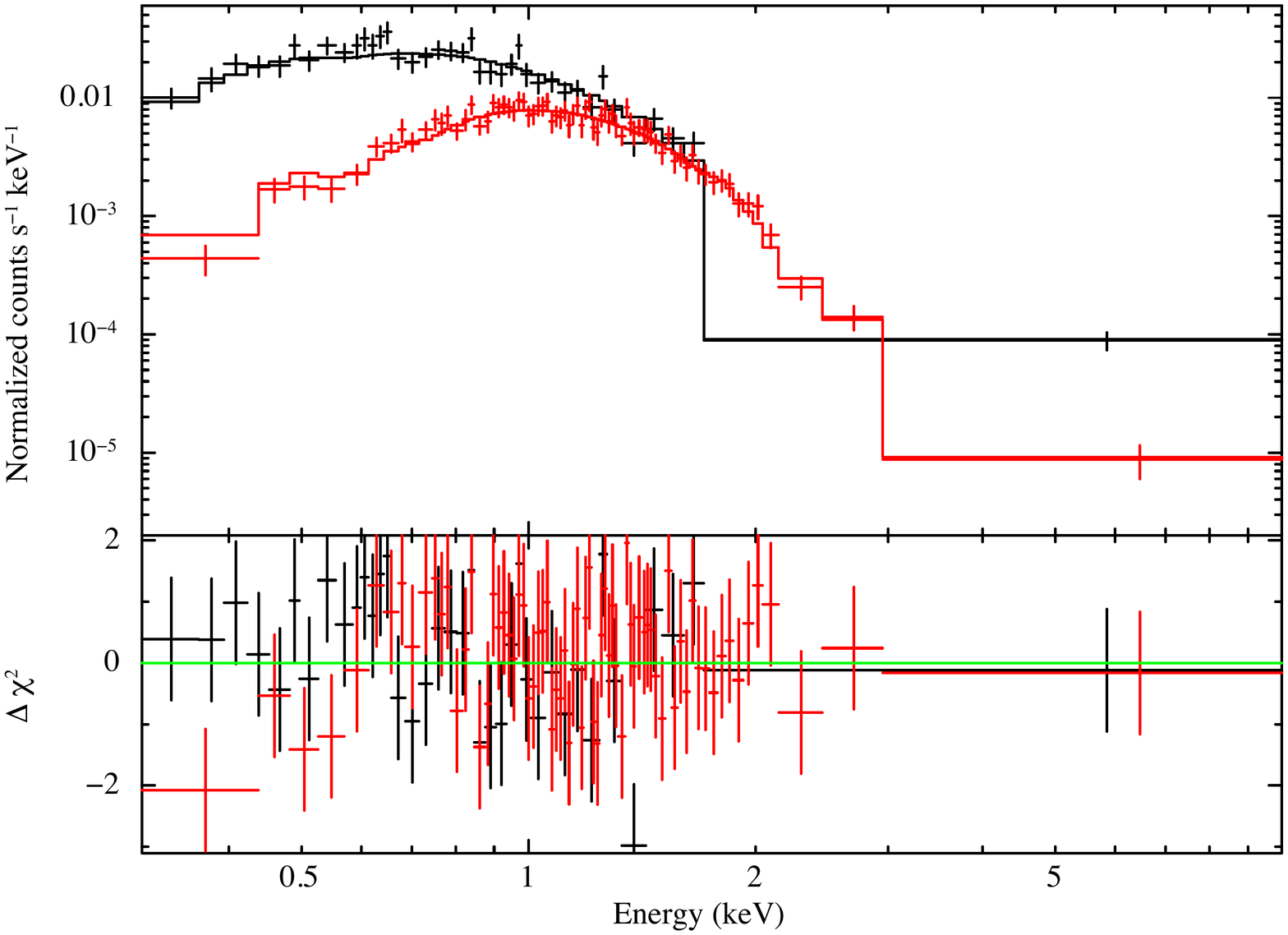}
    \end{subfigure}
    \begin{subfigure}{0.499\textwidth}
        \includegraphics[width=\textwidth]{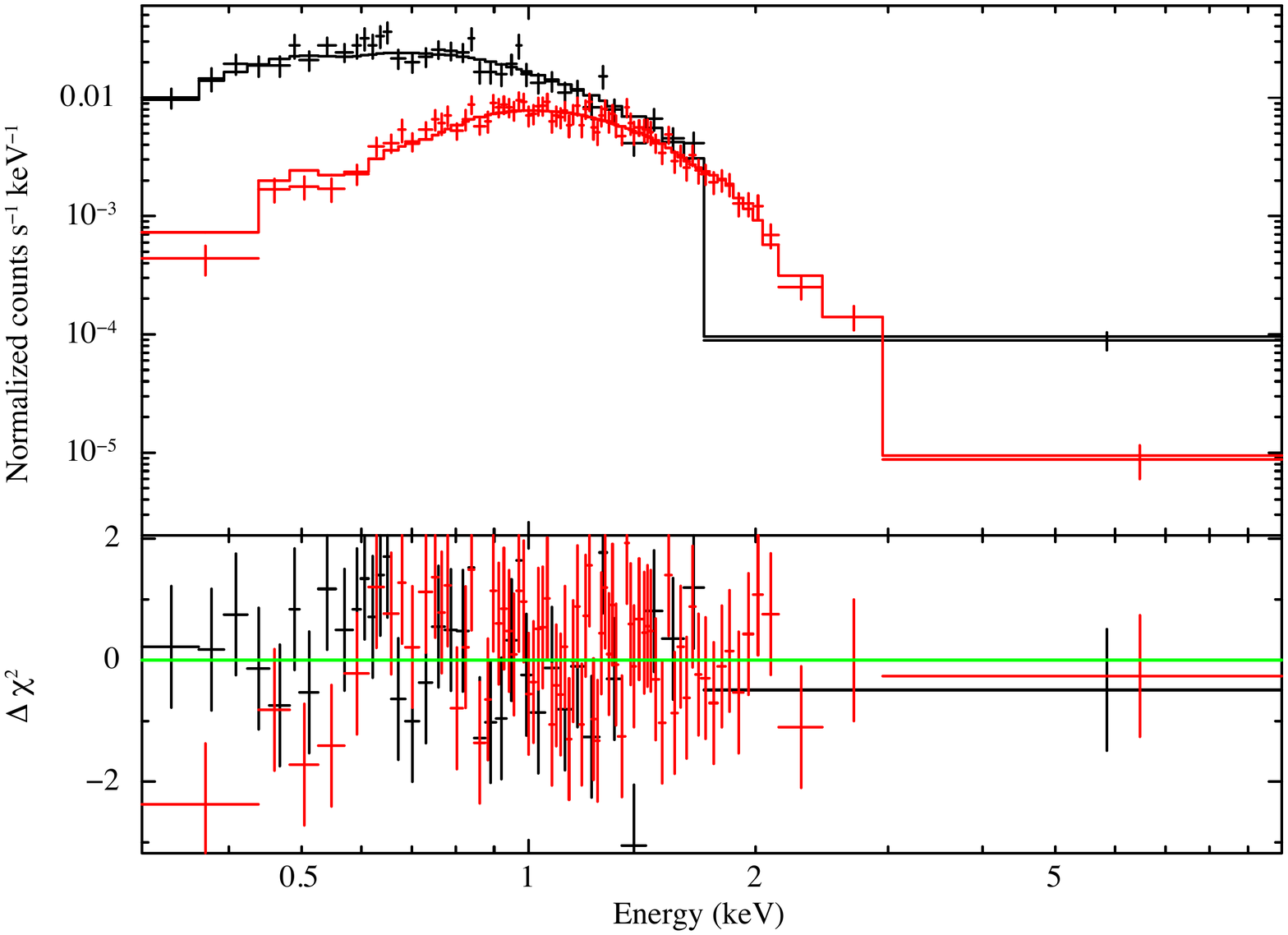}
    \end{subfigure}
    \caption{Spectra of the qLMXB in M30 from the 2001 (black) and 2017 (red) observations, where the reduced sensitivity of \textit{Chandra} in those 16 years can be noticed at low energies. The solid lines represent the best-fit to a \texttt{nsatmos} H-atmosphere (left) and \texttt{nsx} He-atmosphere (right) model. In both cases, a Galactic absorption component (\texttt{tbabs}), a power-law component (with photon index fixed to 1.5) and a pile-up correction model are included. The bottom panels display the differences between data points and the best-fit models.}
    \label{fig:spectra}
\end{figure*}

\begin{figure*}
    \centering
    \begin{subfigure}{0.49\textwidth}
        \includegraphics[width=\textwidth]{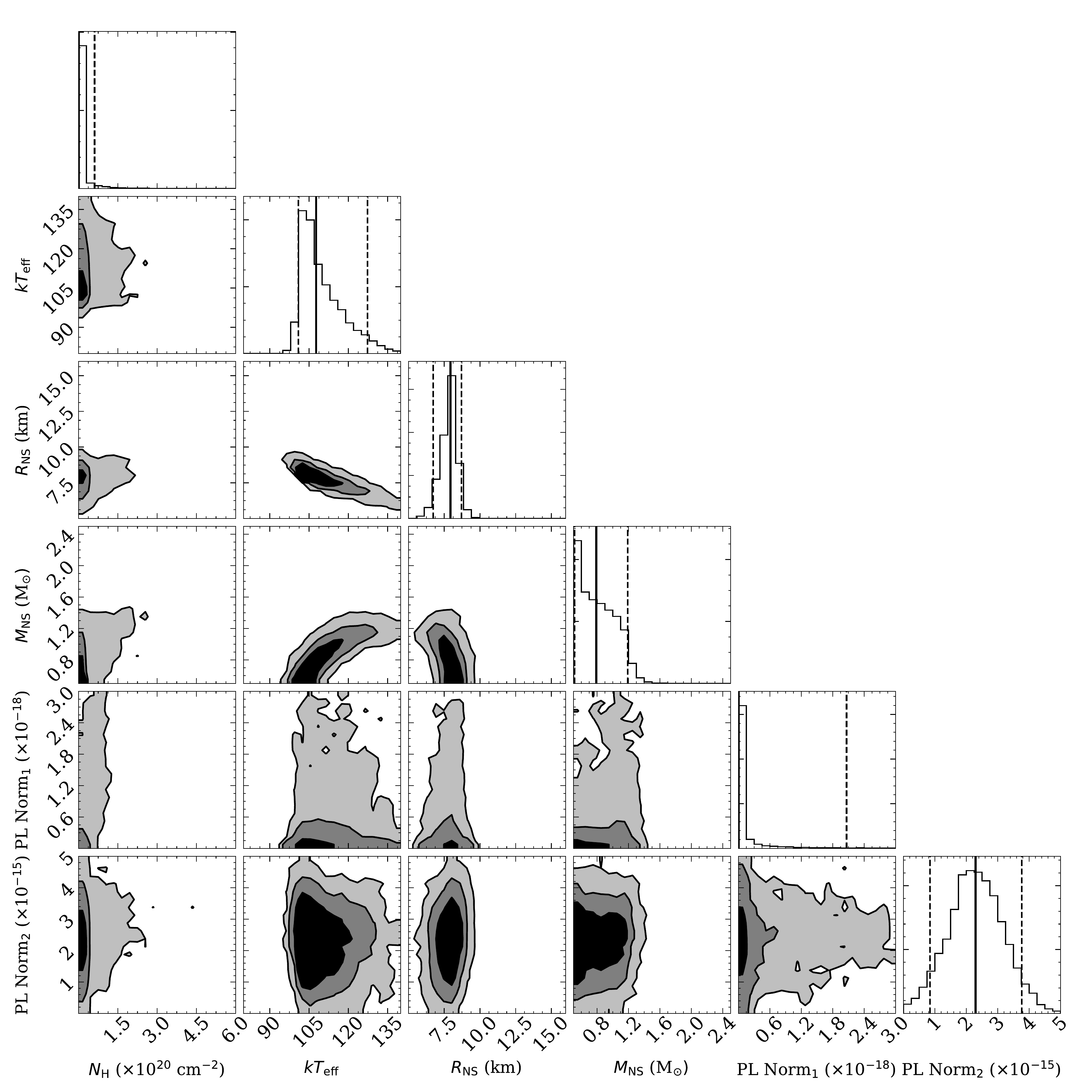}
    \end{subfigure}
    \begin{subfigure}{0.49\textwidth}
        \includegraphics[width=\textwidth]{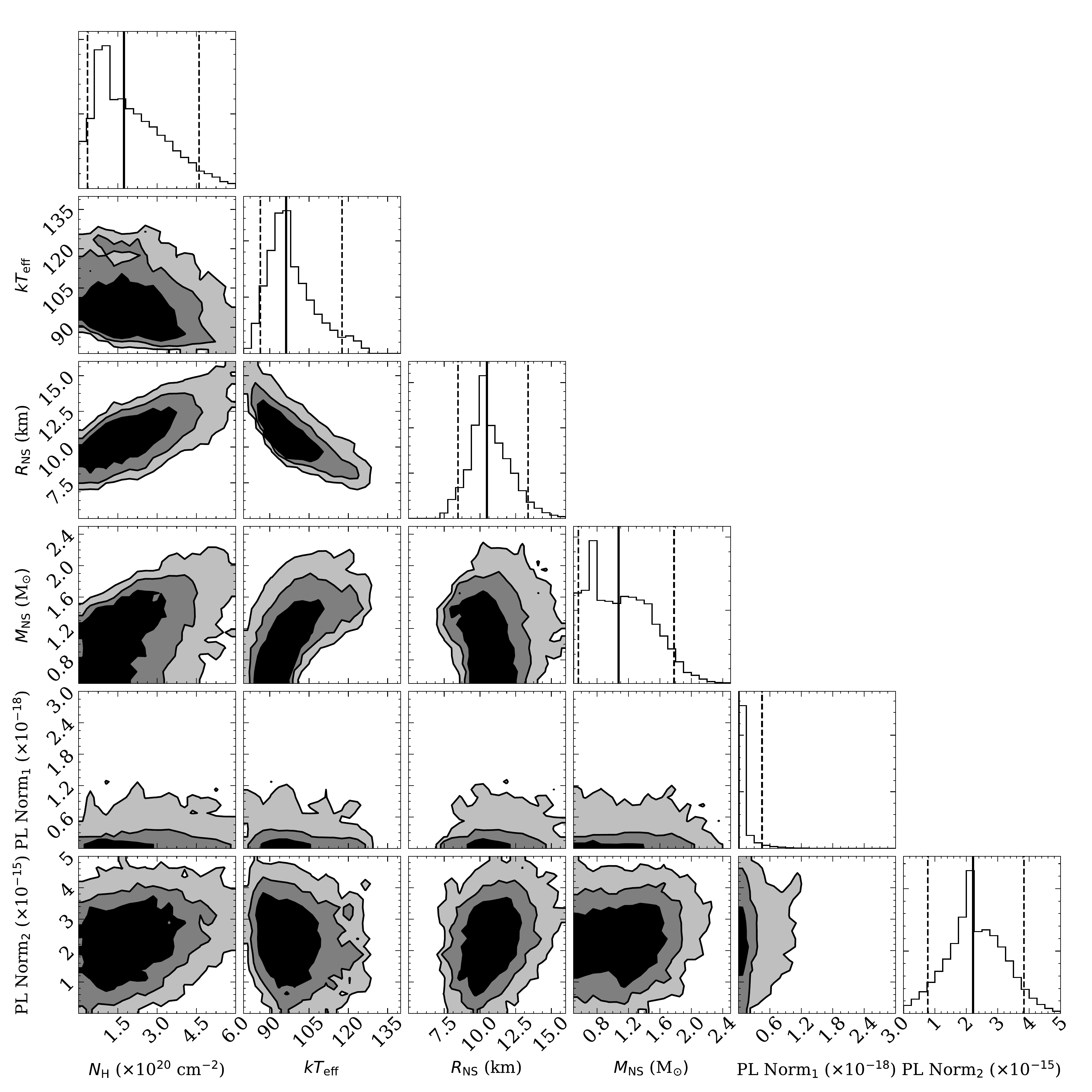}
    \end{subfigure}
    \caption{The one- and two-dimensional MCMC posterior distributions for the parameters in the spectral fitting  (Section~\ref{sec:atmospheres}). The black, dark gray and light gray regions correspond to the 68\%, 90\% and 99\% credible intervals, respectively, for the \texttt{nsatmos} H-atmosphere (left) and \texttt{nsx} He-atmosphere (right) models. The solid line in the one-dimensional posterior distributions represents the median value, while the left/right dashed lines represent the 5\% and 95\% percentiles, respectively. Note that in the mass parameter the lower limit is technically unconstrained, since it is consistent with the hard limit of \texttt{nsatmos} and \texttt{nsx} models. In both figures, the power-law normalisation of group 1 (PL Norm$_1$) and group 2 (PL Norm$_2$) are measured in units of erg~s$^{-1}$~cm$^{-2}$. }
    \label{fig:allmcmc}
\end{figure*}

The spectral fitting is performed by assigning the same set of parameters to both data groups, i.e., keeping the NS temperature, mass and radius tied between the two groups. All the parameters are left free, except for the distance, the H/He normalisation, the photon index and the pile-up parameters mentioned previously. The spectral fit to the model \texttt{pileup*tbabs*(nsatmos+pegpwrlw)} results in a statistically acceptable fit with $\chisqnu=0.96$ with nhp = 0.59 and is shown in Figure \ref{fig:spectra}. From the MCMC posterior distributions, we deduce the median values as well as the 5\% and 95\% quantiles, which give us the 90\% credible intervals of the parameters. All results are summarized in Table~\ref{tab:nsatmos_table}, and the 1- and 2-dimensions posterior distributions are displayed in Figure~\ref{fig:allmcmc}. In particular, we find $\rns=7.94\ud{0.76}{1.21}$~km and $\mns=0.79\ud{0.40}{0.28}$~M$_{\odot}$. It is important to emphasise that the lower limit in the mass parameter is technically unconstrained, since it is consistent with the hard limit of the model. Because the hard limit defines a range for the prior, it likely affects both the lower limit and the median values. For this reason we note that the less affected and most informative value is the upper limit of this parameter. These results are physically unrealistic for an EoS that can support a 2\msun\ NS  \citep[e.g.,][]{2016ARA&A..54..401O}, but consistent with the previously reported range for this source ($7.0\km\leq\rns\leq19.2\km$, at 90\% confidence, \citealt{2007ApJ...657..286L}). Since assumptions of H-atmospheric composition may lead to an underestimation of the NS radius (see Section~\ref{sec:intro}), we next consider an atmosphere composed of pure helium. 

This analysis is performed following the same procedure described before, but with the \texttt{nsx} model\footnote{A "switch" parameter for \texttt{nsx} permits selecting He or carbon composition.} \citep{2009Natur.462...71H}, i.e., \texttt{pileup*tbabs*(nsx+pegpwrlw)}.  The results of the statistically acceptable fit ($\chisqnu=0.98$ with nhp = 0.55) are shown in Table~\ref{tab:nsatmos_table}, Figures~\ref{fig:spectra} and \ref{fig:allmcmc}. Again, we use MCMC simulations to obtain the posterior distributions of the parameters. As expected, we find larger inferred values for \rns\ and \mns: $\rns=10.50\ud{2.88}{2.03}$~km and $\mns=1.07\ud{0.71}{0.51}$~M$_{\odot}$. As explained before, note that the most meaningful value is the upper limit of the mass parameter. These results are easier to reconcile with other measurements of these two NS properties \citep{2016ARA&A..54..401O}, and based on this, one can argue that our data is better described by a He atmosphere model. The best-fit $N_\mathrm{H}$ obtained with  \texttt{nsx} appears consistent with that deduced from the $N_\mathrm{H}$--$E(B-V)$ relations \citep[][for a recent estimate]{foight16}, predicting $N_\mathrm{H}\sim2.8\tee{20}\percmsq{}$, for $E(B-V)=0.03$ \citep{harris96,harris10}. While this consistency is interesting, it is important to keep in mind that $N_\mathrm{H}$--$E(B-V)$ relations may be uncertain, especially for low-absorption. Nonetheless, it is important to note that H and He models describe equally well the data, and are virtually indistinguishable at the signal-to-noise ratio available from current X-ray observatories. 

Additionally, to explore the impact of adding a power-law to both models, we repeated the spectral fits without considering this component.  The spectral fit to the model \texttt{pileup*tbabs*nsatmos} results in $\chisqnu=0.99$ with nhp = 0.49. The measured properties are $\rns=6.51\ud{1.79}{1.40}$~km and $\mns=1.16\ud{0.17}{0.45}$~M$_{\odot}$. While in the case of \texttt{pileup*tbabs*nsx}, the spectral fit results in $\chisqnu=0.99$ with nhp = 0.51, $\rns=8.05\ud{2.49}{1.87}$~km and $\mns=1.39\ud{0.32}{0.58}$~M$_{\odot}$. These results are consistent with those of the previous spectral fits, however, they suggest even smaller $\rns$. Together with the power-law normalisations of 2017 (marginally consistent with zero, see the posterior distributions in Figure~\ref{fig:allmcmc}), this indicates a possible non-negligible contribution of the power-law component to the seemingly purely thermal spectrum. Moreover, if a non-thermal component is indeed present and not accounted for, it may skew the $\rns$ measurement. Because it is appropriate to include the power-law component when trying to place constraints on the radius, we favour our analyses including the power-law.


Using the \texttt{simftest} tool in {\sc xspec}, we also determine the significance of adding an extra power-law component (\texttt{pegpwrlw})
to the \texttt{pileup*tbabs*nsatmos} model. After 1000 iterations with \texttt{simftest}, we find a probability of 0.02 that the data is consistent with the simpler model, i.e., without the power-law component. Likewise, employing \texttt{simftest} to the model \texttt{pileup*tbabs*(nsx+pegpwrlw)} results in a probability of 0.04 that the data is consistent with the model without the \texttt{pegpwrlw} component. Both represent marginal evidence that the power-law component is required. Therefore, we add this component to account for the uncertainties it may cause, as explained previously.


We also repeated all the reported analyses without combining the eight 2017 spectra. The results are shown in Table \ref{tab:indiv_table}. Though we find consistent results, we note the uncertainties are slightly larger, especially for the radius measurements. In particular, when fitting a H atmosphere the upper limit for $\rns$ seems to be consistent with typical NS radii. However, we favour our analyses with combined 2017 spectra, since the precision of measurements is substantially increased.

Finally, due to the reduced sensitivity of \emph{Chandra} at low energies (see Figure \ref{fig:spectra}), we tested that the calibration of the degraded response did not affect the results, and found that the measured radius did not change when using a 6-month old response matrix.

\begin{table}
	\centering
	\caption{Spectral fit parameters for \texttt{nsatmos} and \texttt{nsx}. The statistical fits are \Chisq{0.96}{105}{0.59} for \texttt{nsatmos}, and \Chisq{0.98}{105}{0.55} for \texttt{nsx}. The values in parentheses correspond to fixed parameters in the models, while other values are obtained from the MCMC posterior medians. All reported uncertainties correspond to the 5\% and 95\% quantiles, i.e., representing the 90\% credible intervals.}
	\label{tab:nsatmos_table}
	\begin{threeparttable}
	\begin{tabular}{lcccr} 
		\hline
		Model Parameter & Group 1  & Group 2  \\
		& (2001) & (2017)\\
		\hline
		\multicolumn{3}{c}{\texttt{nsatmos}}\\
		\hline
		Frame time (s) \hspace{1mm}& (3.14)&  (0.941)\\
		\noalign{\vskip 1.5mm}
		$\alpha_{\textrm{pileup}}$ & 0.18\raisebox{0.3ex}{$\scriptstyle^{+0.56}_{-0.17}$}& 0.53\raisebox{0.3ex}{$\scriptstyle^{+0.43}_{-0.48}$} \\
		\noalign{\vskip 1.5mm}  
		$N_\mathrm{H}$ (10$^{20}$ cm$^{-2})$$^{\textrm{a}}$ & \multicolumn{2}{c}{< 0.61} \\
		\noalign{\vskip 1.5mm}  
		$kT{\scriptsize{\textrm{eff}}}$ (eV) & \multicolumn{2}{c}{107.72\raisebox{0.3ex}{$\scriptstyle^{+19.63}_{-6.78}$}} \\
		\noalign{\vskip 1.5mm}  
		$\mns$ (M$_{\odot})^{\textrm{b}}$ & \multicolumn{2}{c}{0.79\raisebox{0.3ex}{$\scriptstyle^{+0.40}_{-0.28}$}} \\
		\noalign{\vskip 1.5mm}  
		$\rns$ (km) & \multicolumn{2}{c}{7.94\raisebox{0.3ex}{$\scriptstyle^{+0.76}_{-1.21}$}} \\
		\noalign{\vskip 1.5mm}  
		$\Gamma^{\textrm{c}}$ & \multicolumn{2}{c}{(1.5)} \\
		\noalign{\vskip 1.5mm}
		PL Norm (10$^{-15}$ erg s$^{-1}$ cm$^{-2})$ & < 0.002$^{\textrm{a}}$& 2.30\raisebox{0.3ex}{$\scriptstyle^{+1.47}_{-1.45}$} \\
		\hline
		\multicolumn{3}{c}{\texttt{nsx}}\\
		\hline
		Frame time (s) \hspace{1mm}& (3.14)&  (0.941) \\
		\noalign{\vskip 1.5mm} 
		$\alpha_{\textrm{pileup}}$ & 0.02\raisebox{0.3ex}{$\scriptstyle^{+0.60}_{-0.02}$}& <0.1$^{\textrm{a}}$ \\
		\noalign{\vskip 1.5mm} 
		$N_\mathrm{H}$ (10$^{20}$ cm$^{-2})$ & \multicolumn{2}{c}{1.74\raisebox{0.3ex}{$\scriptstyle^{+2.87}_{-1.40}$}} \\
		\noalign{\vskip 1.5mm} 
		$kT{\scriptsize{\textrm{eff}}}$ (eV) & \multicolumn{2}{c}{96.28\raisebox{0.3ex}{$\scriptstyle^{+21.30}_{-9.90}$}} \\
		\noalign{\vskip 1.5mm} 
		$\mns$ (M$_{\odot})^{\textrm{b}}$ & \multicolumn{2}{c}{1.07\raisebox{0.3ex}{$\scriptstyle^{+0.71}_{-0.51}$}} \\
		\noalign{\vskip 1.5mm} 
		$\rns$ (km) & \multicolumn{2}{c}{10.50\raisebox{0.3ex}{$\scriptstyle^{+2.88}_{-2.03}$}} \\
		\noalign{\vskip 1.5mm}  
		$\Gamma^{\textrm{c}}$ & \multicolumn{2}{c}{(1.5)} \\
		\noalign{\vskip 1.5mm}
		PL Norm (10$^{-15}$ erg s$^{-1}$ cm$^{-2})$& < 0.0004$^{\textrm{a}}$& 2.21\raisebox{0.3ex}{$\scriptstyle^{+1.62}_{-1.44}$} \\
		\hline
	\end{tabular}
	\begin{tablenotes}
\item[a] Corresponds to the 90\% confidence upper limit.
\item[b] Note that the lower limit of the parameter is unconstrained, and consistent with the hard limit of the model, imposed by the prior. Since the median value may be also affected, the most meaningful value is the upper limit.
\item[c] Represents the photon index of the power-law model.
\end{tablenotes}
\end{threeparttable}
\end{table}

\section{Discussion}\label{sec:discussion}

Our spectral analysis of the qLMXB in the GC M30 seems to favour the He atmosphere model, according to the radius obtained, but it reveals a relatively small radius when using a H model, in apparent disagreement with theoretical and previous experimental studies \citep[see,][for a recent review]{lattimer16}. Specifically, a NS with radius $\rns=7.94\ud{0.76}{1.21}$~km is impossible to reconcile with nucleonic EoS models, especially when those are constrained by NS mass measurements around $\sim2\msun$ \citep{demorest10,antoniadis13,cromartie19}. Such measurements impose a certain stiffness to the EoS which then corresponds to NSs with $\rns\gtrsim 11\km$. Strange quark stars models could however be compatible with the radius measurement obtained with the H atmosphere model \citep{1986A&A...160..121H,2001ApJ...550..426L}.  We nonetheless insist that this is insufficient evidence to claim that this qLMXB in M30 hosts a quark star. The NS radius inferred from qLMXB observations is highly sensitive to the spectral shape, and it is fundamental to explore possible biases induced by the assumptions made in the present analysis. Motivated by this, we discuss below the most relevant and dominant sources of systematic uncertainty.

\subsection{Chemical composition of the atmosphere}

The assumption on the atmospheric composition directly impacts the resulting \mns\ and \rns\ measurements and several studies have shown that the \rinfty\ measured could be underestimated by 20--50\% if a H-atmosphere model is assumed, when a He-atmosphere is actually present atop the NS \citep{2012MNRAS.423.1556S,2013ApJ...764..145C,2014MNRAS.444..443H,2016ApJ...831..184B,2018MNRAS.476..421S}. In these works and in the present one, both H and He atmosphere models fit the data equally well. Spectral analyses alone cannot distinguish between the two possible compositions.

As stated before, while the atmospheric composition of a NS in a qLMXB is expected to be pure H, it depends on the nature of the donor star. In UCXBs, the companion star could be a helium star or a white dwarf. In fact, some of them are known to possess a helium-rich donor star \citep{2010ApJ...712L..58A,2017A&A...598A..34S,2019ApJ...875...25C}, or to exhibit He-ignition bursts from the NS surface indicating accretion of He \citep[see][for a review]{Strohmayer06}.

Identifying the presence of hydrogen in a LMXB system, or understanding the nature of the donor is key to determine the atmosphere composition of the NS. This can be achieved by the detection of a \halpha\ emission line in its optical spectrum, which would indicate the presence of a hydrogen-rich accretion disk  \citep[see][for the qLMXB in \OmCen]{2004ApJ...613..512H}. If H is transferred onto the NS, then the surface (i.e., the atmosphere) must necessarily be composed of hydrogen. Alternatively, evidence of H composition may be proved by measuring the orbital period of the binary. As in the case of the qLMXB X5 in 47~Tuc, the 8.666$\pm$0.008 hour eclipse period allowed to constrain the properties of the companion, indicating it must be a main-sequence star \citep{2002ASPC..271..349H}.

For the qLMXB in M30, the orbital period is unknown, no information is available on the nature of the companion star or on the presence of H in the binary system.  These imply that the chemical composition of the NS atmosphere remains uncertain, and a He composition cannot be excluded.

Even if the donor star is H-rich, another process that may alter the chemical composition of a NS atmosphere is the nuclear diffuse burning of H into He, where an underlying layer can consume H by capturing protons on timescales of $10^4-10^5$ yr \citep{2003ApJ...585..464C,2004ApJ...605..830C}. The H depletion implies a heavier element composition of the external atmosphere \citep[e.g.,][]{2010ApJ...723..719C}, favoring a He composition. If that is the case, and given the large timescale of the process, another source of heating is required to account for the $\sim10^{33}$~erg~s$^{-1}$ luminosity. An example of this are the new class of NSs known as `hot widows', thought to be descendants of NSs in LMXBs, which are powered by r-mode instability, capable of maintaining high temperatures during $\sim10^9$ yr \citep{2014MNRAS.445..385C}. The nuclear diffusive burning scenario requires that no continuing accretion is occurring, as it would quickly replenish the atmosphere with H. However, the presence of the variable power-law component strongly suggests there is continuing accretion. Otherwise, it could represent a propeller shock, meaning that the material does not land on the NS surface, and allowing the nuclear diffusive burning hypothesis.

\subsection{Surface temperature inhomogeneities}
\label{subsec:surf}

To model the emission from the NS, we assumed that the entire surface emits uniformly. However, unmodeled surface inhomogeneities, e.g., surface hot spots, could bias the measured radius. The presence of hot spots on the surface of the NS could be detectable as pulsed emission depending on the circumstances, e.g., the size, temperature and location of the spots, or the inclination of the observer with respect to the rotation axis \citep{2016ApJ...826..162E}. To detect such pulsations an instrument with high timing resolution and high sensitivity is required. Currently, there are no X-ray observations of qLMXBs with sufficient timing resolution, except for deep \emph{Chandra} High Resolution Camera observations of 47~Tuc \citep{2007ApJ...660..587C}, and M28 \citep{2004ApJ...613..522R}.

Ignoring a possible surface hot spot when modeling the emission spectrum would result in a measured temperature higher than the true value, and therefore a radius smaller than the actual NS value. \cite{2016ApJ...826..162E} carefully studied the biases caused by unmodeled hot spots on the surface of NSs in qLMXBs; basing their calculations on the upper limit on the pulsed fraction for the qLMXB X7 in 47~Tuc, $\lesssim 16\%$, obtained from the \chandra-HRC observations.  They determined that, in the case of 47~Tuc~X7, not modeling hot spots that might be present on the NS surface would cause an underestimation of the radius by as much as $\sim 25\%$ \citep{2016ApJ...826..162E}. In this context, we tested the presence of a hot spot by fitting the model \texttt{pileup*tbabs*(nsatmos+pegpwrlw)} with \texttt{nsatmos} normalisation free to vary, a fixed radius $\rns=12$~km, and power-law photon index $\Gamma=1.5$. The result of the spectral fit is an emitting area of $38\ud{9}{6}$\% of the total NS surface, where the uncertainty corresponds to the 90\% confidence computed with \texttt{error} command. Since typically hot spot sizes represent less than 2\% of the surface emitting area \citep[e.g.,][]{2016MNRAS.463.2612G}, we conclude the area given by the fit is probably larger than a hot spot. However, it is consistent with a spectrum that combines the emission from the full surface and a small hot spot.

The present \chandra-ACIS observations of M30 lack the time resolution (and the sensitivity) to search for the existence of hot spots. However, they highly motivate searches for X-ray pulsations in the future. The {\it Neutron Star Interior Composition Explorer} (\nicer, \citealt{gendreau17}) could address this issue \citep{2018IAUS..337..187R}, but the non-imaging capabilities will somewhat complicate the search for pulsations in the crowded field of GCs. In the next decade, the {\it Athena X-ray Observatory} \citep{nandra13} will be able to provide X-ray data with high sensitivity, $10\,\mu{\rm s}$ time resolution and $\sim$5\arcsec\ spatial resolution, that will facilitate the search for the presence of X-ray pulsations as evidence of hot spots at the surface of NSs in qLMXBs.

\subsection{Rotational corrections}
In this work we have made the assumption that the NS is spherically symmetric and is not rotating. However, some NSs rotate at high frequencies extending up to 716 Hz \citep{2006Sci...311.1901H}, and implying surface velocities $\gtrsim 0.1c$, as revealed from their radio pulsations. Other NSs spin frequencies have been discovered thanks to their X-ray pulsations, as the case of accreting millisecond X-ray pulsars, rotating hundreds of times per second \citep[see][for a review]{2012arXiv1206.2727P}.

The effect of fast rotation is to distort the spectrum with respect to a non-rotating one, because photons experience a Doppler shift when they are emitted from the approaching or receding side of the stellar surface. Also, such rapid spins distort the shape of the surface, which becomes oblate \citep{2014ApJ...791...78A}. Thus, the fast rotation of NSs may impact the observed thermal spectrum, and consequently, bias the measurements of the spectroscopic radius $R_{\infty}$ by a few percent.
\cite{2015ApJ...799...22B} quantified the bias in the inferred $R_{\infty}$ when the non-spinning case is assumed compared to a spinning case, from moderate to fast rotation. 

The corrections to the radius $R_{\infty}$ depend on the NS spin, but also on its compactness (i.e., on \mns\ and \rns) and on the observer's viewing angle with respect to the rotation axis.  Not considering these corrections results in an underestimation of the flux by as much as 12\%, the worst case scenario being for an edge-on fast rotating NS ($\sim$600\,Hz) with a 15\km\ radius. \cite{2015ApJ...799...22B} estimated the incurred bias on the measured NS radius which can be as much as 10\% for a 800\,Hz, 15\km\,NS.  For more realistic NS properties (e.g., a 11\km\ NS spinning at 600\,Hz), the radius could be underestimated by $\sim2.5\%$ if the rotational correction are neglected. Note that these corrections were calculated for the Blackbody case, which may be different from the H-atmosphere case (Amason \& Morsink, in preparation). 

Unfortunately, for the NS in the M30 qLMXB, like for most qLMXBs in GCs, the spin frequency is unknown, since these objects do not exhibit radio emission, and their (likely) uniform surface temperature precludes the detection of pulsations in the X-ray band. However, if surface inhomogeneities are present, future missions will be capable of detecting them, as discussed in section \ref{subsec:surf}.

\subsection{Distance}
Another source of uncertainty in the radius measurements is the distance $d$ to the GC hosting the qLMXB, since $F\sim (R_{\infty}/d)^2$.  We considered the most recent distance measurement to M30, 8.2\ppm0.62\kpc\ \citep[$1\sigma$,][]{2017ApJ...838..162O}, determined with the main sequence (MS) fitting method, which consists in matching the MS stars of the GC to a theoretical main sequence in a colour-magnitude diagram (CMD). The latter is obtained from stellar models, which require calibrations using stars with known distances. The GC main sequence is shifted until it matches the theoretical one, and that shift is related to the \emph{distance modulus}, allowing to estimate the distance. The reported value of 8.2\ppm0.62\kpc\ incorporates uncertainties associated to the theoretical model, the photometric measurements, metallicity and reddening estimates, which could affect the location of the stars in the CMD. 

The uncertainty in the source distance adds to the uncertainty on \rinfty, and therefore on \rns. In the case of M30, the distance uncertainty of \ppm7.6\% (at 1$\sigma$ confidence) is not dominant compared to the other sources of uncertainties discussed above, such as the chemical composition of the atmosphere, or surface temperature inhomogeneities. Moreover, it is smaller than the statistical uncertainties on the radius in our results. Thus, the uncertainty on $d$ was not included in our analysis. In the near  future, precise measurements of the distance to GCs are expected to be achievable with the next \gaia\ data release, permitting distances with uncertainties better than 1\%  \citep{2017MNRAS.467..412P}.

\section{Conclusions}\label{sec:conclusion}
In this work, we have performed a spectral analysis of two archived \emph{Chandra} observations of the qLMXB, CXOGLB~J214022.13$-$231045.5, in the GC M30. Because the two observations are separated by 16 years, we first confirmed that  the flux remained constant at 90\% confidence. We then performed a simultaneous spectral analysis of the two data sets using a model commonly used for qLMXB: a light-element composition atmosphere (H/He), absorbed by interstellar medium, corrected for pile-up, and a power-law for count excesses at high energies. Our main conclusions are the following:
\begin{enumerate}
\item[1.] 
    The spectral fit performed with the H-atmosphere model \texttt{nsatmos}  results in relatively small radii, but consistent, compared to the previous results reported by \cite{2007ApJ...657..286L}. The MCMC simulations we performed to quantify the parameters' credible regions give a measured radius $\rns=7.94\ud{0.76}{1.21}$~km and a mass $\mns<1.19$~M$_{\odot}$.
\item[2.] 
    When considering the He-atmosphere model \texttt{nsx}, the measured radius and mass are somewhat larger, as expected, $\rns=10.50\ud{2.88}{2.03}$~km and $\mns<1.78$~M$_{\odot}$, but consistent with typically measured values for qLMXBs and other classes of NSs \citep{2016ARA&A..54..401O}. However, it is worth to mention that the error range is large enough to include most of both the observed and predicted ranges. Because this spectral model results in NS properties that are less at odds with previous measurements, we tentatively conclude that it provides a more appropriate description of the qLMXB in M30.
\item[3.] 
    We find that a variable power-law component is required in addition to the thermal component (\texttt{nsatmos} or \texttt{nsx}), to properly constrain the $\rns$ measurement.
\end{enumerate}

Identifying the companion star, or verifying the presence of H or He in the qLMXB system, will determine the spectral model to use for the X-ray spectrum of this source. Other sources of uncertainties and biases discussed in Section~\ref{sec:discussion}, in particular those that cause an underestimation of the radius (surface temperature inhomogeneities, being the strongest source of bias, since it accounts for $\sim25$\% of uncertainty) could possibly make the H radius reported above more compatible with NS radii in the  11--14\km\ range that have been reported recently \citep[e.g.,][]{nattila17,abbott18,baillot19,gonzalez19,2019ApJ...887L..22R,2019arXiv191211031R}.

\section*{Acknowledgements}

C.S.E. acknowledges the support of the FONDECYT Regular Project 1150411 during the early stages of this work, the funding from McGill's Wolfe Fellowship in Scientific and Technological Literacy, the McGill Space Institute, and the Canadian Institute for Advanced Research (CIFAR). C.S.E. also thanks Daryl Haggard for valuable discussions. S.G. and N.W. acknowledge the support of the French Centre National d'\'{E}tudes Spatiales (CNES). C.O.H. is supported by NSERC Discovery Grant RGPIN-2016-04602, and a Discovery Accelerator Supplement. We also acknowledge extensive use of NASA's Astrophysics Data System (ADS) Bibliographic Services and the ArXiv.  Finally, we acknowledge the Chandra scheduling, data processing, and archive teams for making these observations possible.


\bibliographystyle{mnras}
\bibliography{mnras_template}

\appendix
\section{} \label{sec:appendix2}
\begin{table}
	\centering
	\caption{Spectral fit parameters for \texttt{nsatmos} and \texttt{nsx}, for individual 2017 spectra. The statistical fits are \Chisq{0.96}{125}{0.62} for \texttt{nsatmos}, and \Chisq{0.95}{125}{0.63} for \texttt{nsx}. The values in parentheses correspond to fixed parameters in the models, while other values are obtained from the MCMC posterior medians. All reported uncertainties correspond to the 5\% and 95\% quantiles, i.e., representing the 90\% credible intervals.}
	\label{tab:indiv_table}
	\begin{threeparttable}
	\begin{tabular}{lcccr} 
		\hline
		Model Parameter & Group 1  & Group 2  \\
		& (2001) & (2017)\\
		\hline
		\multicolumn{3}{c}{\texttt{nsatmos}}\\
		\hline
		Frame time (s) \hspace{1mm}& (3.14)&  (0.941)\\
		\noalign{\vskip 1.5mm}
		$\alpha_{\textrm{pileup}}$ & 0.09\raisebox{0.3ex}{$\scriptstyle^{+0.51}_{-0.09}$} & $< 0.78^{\textrm{a}}$ \\
		\noalign{\vskip 1.5mm}  
		$N_\mathrm{H}$ (10$^{20}$ cm$^{-2})$ & \multicolumn{2}{c}{1.20\raisebox{0.3ex}{$\scriptstyle^{+2.72}_{-1.09}$}} \\
		\noalign{\vskip 1.5mm}  
		$kT{\scriptsize{\textrm{eff}}}$ (eV) & \multicolumn{2}{c}{106.46\raisebox{0.3ex}{$\scriptstyle^{+29.97}_{-14.19}$}} \\
		\noalign{\vskip 1.5mm}  
		$\mns$ (M$_{\odot})$ & \multicolumn{2}{c}{1.06\raisebox{0.3ex}{$\scriptstyle^{+0.50}_{-0.51}$}} \\
		\noalign{\vskip 1.5mm}  
		$\rns$ (km) & \multicolumn{2}{c}{8.63\raisebox{0.3ex}{$\scriptstyle^{+2.75}_{-1.98}$}} \\
		\noalign{\vskip 1.5mm}  
		$\Gamma^{\textrm{b}}$ & \multicolumn{2}{c}{(1.5)} \\
		\noalign{\vskip 1.5mm}
		PL Norm (10$^{-15}$ erg s$^{-1}$ cm$^{-2})^{\textrm{a}}$ & < 0.005 & <0.008 \\
		\hline
		\multicolumn{3}{c}{\texttt{nsx}}\\
		\hline
		Frame time (s) \hspace{1mm}& (3.14)&  (0.941) \\
		\noalign{\vskip 1.5mm} 
		$\alpha_{\textrm{pileup}}^{\textrm{a}}$ & $<0.001$ & $<2.72\times 10^{-5}$ \\
		\noalign{\vskip 1.5mm} 
		$N_\mathrm{H}$ (10$^{20}$ cm$^{-2})$ & \multicolumn{2}{c}{2.24\raisebox{0.3ex}{$\scriptstyle^{+2.88}_{-1.90}$}} \\
		\noalign{\vskip 1.5mm} 
		$kT{\scriptsize{\textrm{eff}}}$ (eV) & \multicolumn{2}{c}{94.35\raisebox{0.3ex}{$\scriptstyle^{+20.24}_{-11.42}$}} \\
		\noalign{\vskip 1.5mm} 
		$\mns$ (M$_{\odot})$ & \multicolumn{2}{c}{1.19\raisebox{0.3ex}{$\scriptstyle^{+0.74}_{-0.60}$}} \\
		\noalign{\vskip 1.5mm} 
		$\rns$ (km) & \multicolumn{2}{c}{11.30\raisebox{0.3ex}{$\scriptstyle^{+3.59}_{-2.42}$}} \\
		\noalign{\vskip 1.5mm}  
		$\Gamma^{\textrm{b}}$ & \multicolumn{2}{c}{(1.5)} \\
		\noalign{\vskip 1.5mm}
		PL Norm ($10^{-15}$ erg s$^{-1}$ cm$^{-2}$)$^{\textrm{a}}$ & < 0.006 & <0.007 \\
		\hline
	\end{tabular}
	\begin{tablenotes}
\item[a] Corresponds to the 90\% confidence upper limit.
\item[b] Represents the photon index of the power-law model.
\end{tablenotes}
\end{threeparttable}
\end{table}

\begin{table}
 \caption{The best-fit fluxes for individual 2017 observations estimated from the MCMC posterior medians, where uncertainties correspond to the 5\% and 95\% quantiles.}
 \label{tab:indv_fluxes}
 \begin{tabular}{lcr}
  \hline
  ObsID & $\log_{10}(F/\cgsflux)$ \\
  \hline
  18997 & -13.14\ud{0.05}{0.05} \\
  20725 & -13.23\ud{0.10}{0.12} \\
  20726 & -13.12\ud{0.09}{0.12} \\
  20731 & -13.08\ud{0.08}{0.10} \\
  20732 & -13.12\ud{0.06}{0.06} \\
  20792 & -13.10\ud{0.07}{0.08} \\
  20795 & -13.25\ud{0.12}{0.16} \\
  20796 & -13.09\ud{0.07}{0.08} \\
  \hline
 \end{tabular}
\end{table}


\bsp	
\label{lastpage}
\end{document}